\documentclass[aps,preprint,floats,epsf,epsfig,nofootinbib,letter]{revtex4}
\usepackage{epsfig}
\usepackage{graphicx}% Include figure files
\usepackage{dcolumn}% Align table columns on decimal point
\usepackage{bm}% bold math
\usepackage{subfigure}

\def\DESepsf(#1 width #2){\epsfxsize=#2 \epsfbox{#1}}
\begin{document}
\def\be{\begin{eqnarray}}
\def\en{\end{eqnarray}}
\def\non{\nonumber}
\def\la{\langle}
\def\ra{\rangle}
\def\Br{{\mathcal B}}
\def\BB{{{\cal B} \overline {\cal B}}}
\def\BD{{{\cal B} \overline {\cal D}}}
\def\DB{{{\cal D} \overline {\cal B}}}
\def\DD{{{\cal D} \overline {\cal D}}}
\def\sq{\sqrt}
\def\ov{\overline}

%\preprint{APS/123-QED}

\title{Bottom Baryon Decays to Pseudoscalar Meson and Pentaquark}
% Force line breaks with \\

\author{Hai-Yang Cheng$^{1}$, Chun-Khiang Chua$^{2}$}
%%\altaffiliation[Also at ]{Physics Department, XYZ University.}
%%Lines break automatically or can be forced with \\
%%\author{}%
%%\email{Second.Author@institution.edu}
\affiliation{$^1$ Institute of Physics, Academia Sinica,
Taipei, Taiwan 115, Republic of China,
$^2$ Department of Physics and Center for High Energy Physics,
Chung Yuan Christian University,
Chung-Li, Taiwan 320, Republic of China}

\date{\today}% It is always \today, today,

\begin{abstract}

Based on SU(3) flavor symmetry, we decompose the decay amplitudes of bottom baryon decays to a pseudoscalar meson and an octet (a decuplet) pentaquark in terms of three (two) invariant amplitudes $T_1$ and $T_{2,3}$ ($\tilde T_1$ and $\tilde T_2$) corresponding to external $W$-emission and internal $W$-emission diagrams, respectively. For antitriplet bottom baryons $\Lambda_b^0,\Xi_b^0$ and $\Xi_b^-$, their decays to a decuplet pentaquark proceed only through the internal $W$-emission diagram. Assuming the dominance from the external $W$-emission amplitudes, we present an estimate of the decay rates relative to $\Lambda_b^0\to P_p^+K^-$, where $P_p^+$ is the hidden-charm pentaquark with the same light quark content as the proton.  Hence, our numerical results will provide a very useful guideline to the experimental search for pentaquarks in bottom baryon decays. For example, $\Xi_b^0\to P_{\Sigma^+}K^-$, $\Xi_b^-\to P_{\Sigma^-}\bar K^0$, $\Omega_b^-\to P_{\Xi^-}\bar K^0$ and $\Omega_b^-\to P_{\Xi^0}K^-$ may have rates comparable to that of $\Lambda_b^0\to P_p^+K^-$ and these modes should be given the higher priority in the experimental searches for pentaquarks.

\end{abstract}

  %Bottom mesons

\maketitle

%%%%%%%%%%%%%%%%%%%%%%%
\section{Introduction}

The LHCb Collaboration has recently announced two hidden-charm pentaquark-like resonances $P_c(4380)^+$ and $P_c(4450)^+$
\footnote{Starting from the next section and thereafter we will use $P_{\cal B}$ to denote the hidden-charm pentaquark with  the same light quark content as the octet or decuplet baryon ${\cal B}$. Hence,  $P_c^+$ with the $\bar ccuud$ quark content will be denoted by $P_p^+$ in our notation.}
in the $J/\psi p$ invariant mass spectrum through the $\Lambda_b^0\to J/\psi pK^-$ decay \cite{LHCb:penta}. The measured masses and widths are:
\be \label{eq:mass}
&& M=4380\pm8\pm29~{\rm MeV}, \qquad\quad \Gamma=205\pm18\pm86 ~{\rm MeV}, ~~{\rm for}~~P_c(4380)^+,
\non \\
&& M=4449.8\pm1.7\pm2.5~{\rm MeV}, \quad ~\Gamma=39\pm5\pm19~ {\rm MeV},
~~~~\,{\rm for}~~P_c(4450)^+.
\en
The best fit solution has spin-parity $J^P$ values of $(3/2^-,5/2^+)$, though acceptable solutions are also found for additional cases with opposite parity, either $(3/2^+,5/2^-)$ or $(5/2^+,3/2^-)$. LHCb has also reported the branching fractions of $\Lambda_b^0\to P_c^+(\to J/\psi p)K^-$ to be \cite{LHCb:Br}
\be \label{eq:BR}
\Br(\Lambda_b^0\to P_c^+K^-)\Br(P_c^+\to J/\psi p)=\cases{ (2.56\pm0.22\pm1.28^{+0.46}_{-0.36})\times 10^{-5} & for~~$P_c(4380)^+$, \cr
(1.25\pm0.15\pm0.33^{+0.22}_{-0.18})\times 10^{-5} & for~~$P_c(4450)^+$. \cr}
\en

The valence quark content of the pentaquark-like resonance is $\bar ccuud$. If this new resonance is indeed a genuine pentaquark state, it is natural to ask what is its nature, such as spin-parity quantum numbers, mass and the internal structure, and what are the dynamical properties, such as strong and weak decays. Many models have been proposed recently to explain the hidden-charm pentaquarks, including (i) a cluster structure for quarks inside the pentaquark, for example, two colored diquarks bound with an antiquark \cite{Maiani:2015vwa,Anisovich:2015cia,Wang:2015epa}, a model  originally proposed by Jaffe and Wilczek \cite{Jaffe:2003sg}, or one diquark and one triquark \cite{Lebed:2015tna} as originally advocated by Karliner and Lipkin \cite{Karliner:2004qw}, (ii) the charmed meson-charmed baryon molecular state,  for example, $P_c(4380)^+$ and $P_c(4450)^+$ being $\bar D\Sigma_c^*$ and $\bar D^*\Sigma_c$ molecular states, respectively \cite{Chen:2015loa,Chen:2015moa,He:2015cea,Roca:2015dva,Burns:2015dwa},
\footnote{Because of their opposite parities, $P_c(4380)$ and $P_c(4450)$ cannot be both the $S$-wave states of $\bar D\Sigma_c^*$ and $\bar D^*\Sigma_c$, respectively. The assignment is opposite in \cite{Chen:2015moa} where $P_c(4380)$ is identified with $\bar D^*\Sigma_c$ and $P_c(4450)$ with the admixture of $\bar D\Sigma_c^*$ and $\bar D^*\Lambda_c$. }
(iii) a composite $\chi_{c1}p$ state for $P_c(4450)^+$  \cite{Meissner:2015mza}, (iv) composite $J/\psi N(1440)$ and $J/\psi N(1550)$ states for $P_c(4380)$ and $P_c(4450)$ \cite{Kubarovsky:2015aaa}, respectively, (v) soliton states for pentaquarks \cite{Scoccola:2015nia}, and (vi)
threshold enhancement or kinematic effect \cite{Guo:2015umn,Liu:2015fea,Mikhasenko:2015vca,Anisovich:2015cia,Anisovich}.

If the pentaquark resonances discovered by the LHCb in $\Lambda_b^0\to J\psi pK^-$ are genuine states, it will be quite important to search for them in other bottom baryon decays, in inclusive production at LHC and $e^+e^-$ factories and in  photoproduction off a proton target \cite{Wang:2015jsa,Kubarovsky:2015aaa,Karliner:2015voa}. Since
LHC can produce a huge number of bottom baryons in addition to $\Lambda_b$'s, it can provide a rich source for the pentaquark production in bottom baryon decays. Under SU(3) symmetry the pentaquark state can be in the octet or decuplet representation. Under a plausible assumption on the relative importance of decay amplitudes, we give an estimate on the decay rates relative to $\Lambda_b^0\to P_c^+K^-$. Hence, our numerical results will provide a very useful guideline to the experimental search for pentaquarks in bottom baryon decays.

This work is organized as follows. In Sec. II we set up the formulism. Under SU(3) flavor symmetry, the bottom baryon decays to a pseudoscalar meson and an octet or a decuplet pentaquark can be expressed in terms of three invariant amplitudes which correspond to two different types of $W$-emission diagrams. Assuming the dominance of one of the $W$-emission amplitude, we proceed to show in Sec. III the numerical estimates for the decay rates relative to $\Lambda_b^0\to P_c^+K^-$. Sec. IV gives our conclusions.

\section{Formalism}

%The effective weak Hamiltonian for charmless $B$ decays is~\cite{Buras}
%\be
%H_{\rm eff}=\frac{G_f}{\sq2}
%                   \Big\{\sum_{r=u,c} V_{qb}V^*_{uq}[c_1 O^r_1+c_2 O^r_2]
%                         -V_{tb} V^*_{tq}\sum_{i=3}^{10} c_i O_i\Big\}+{\rm H.c.},
%\label{eq:H_eff1}
%\en
%where $q=d,s$, and
%\be
%&& O^r_1=(\bar r_\alpha b_\alpha)_{V-A}(\bar q_\beta r_\beta)_{V-A},
%\qquad\qquad
%O^r_2=(\bar r_\beta b_\alpha)_{V-A}(\bar q_\alpha r_\beta)_{V-A},
%\non\\
%&& O_{3(5)}=(\bar q b)_{V-A}\sum_{q'}(\bar q' q')_{V\mp A},
%\qquad\quad
%O_{4(6)}=(\bar q_\alpha b_\beta)_{V-A}\sum_{q'}(\bar q_\beta' q_\alpha')_{V\mp A},
%\non\\
%&& O_{7(9)}=\frac{3}{2}(\bar q b)_{V-A}\sum_{q'}e_{q'}(\bar q' q')_{V\pm A},
%\quad
%O_{8(10)}=\frac{3}{2}(\bar q_\alpha b_\beta)_{V-A}\sum_{q'}e_{q'}(\bar q_\beta' q_\alpha')_{V\pm A},
%\label{eq:H_eff2}
%\en
%with $O_{3-6}$ being the QCD penguin operators, $O_{7-10}$ the electroweak penguin operators, and $(\bar q' q)_{V\pm %A}\equiv \bar q'\gamma_\mu(1\pm\gamma_5)q$.
%

The flavor structure of the weak Hamiltonian governing a weak $\Delta S=-1$ decay at tree level is expressed as
\be
{\cal O}_T\sim (\bar c b )(\bar s u )
 =H^{i} (\bar c b) (\bar q_i c),
\quad
H^{i}=\delta_{i3},
\en
where $\bar q_{1,2,3}=\bar u,\bar d,\bar s$, respectively, and $H^i$ is a spurion field.
The above expression is also applicable to the $\Delta S=0$
case with the $s$ quark field replaced by the $d$ quark one, and with the spurion field defined as $H^{i}=\delta_{i2}$.

%%%%%%%%%%%%%%%%%%%%%%%%%%%%%%%%%%%%%%%%%%%%%%%%%%%%%
%Fig 1%
%%%%%%%%%%%%%%%%%%%%%%%%%%%%%%%%%%%%%%%%%%%%%%%%
\begin{figure}[t]
\centering
 \subfigure[]{
     \includegraphics[width=0.5\textwidth,natwidth=610,natheight=642]{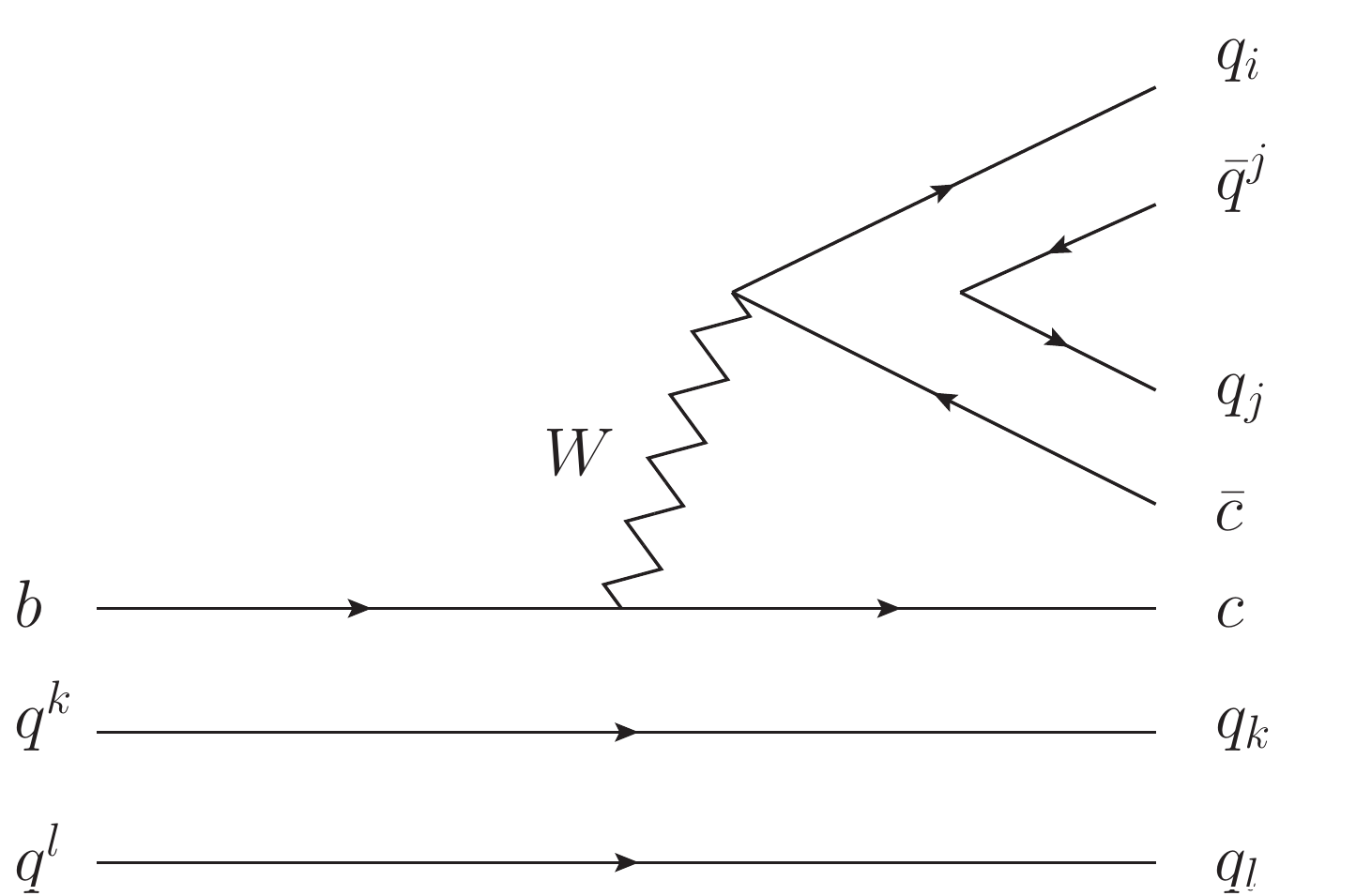}
}\subfigure[]{
    \includegraphics[width=0.5\textwidth,natwidth=610,natheight=642]{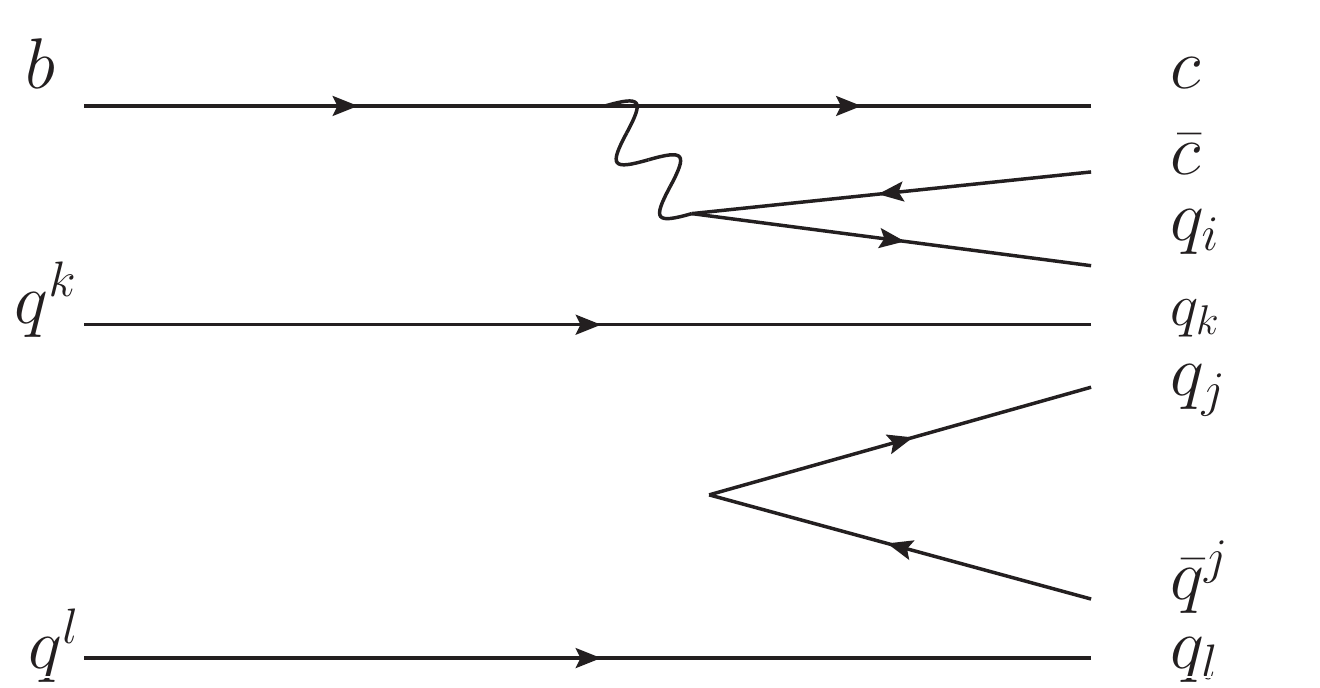}
}\\
\subfigure[]{
    \includegraphics[width=0.5\textwidth,natwidth=610,natheight=642]{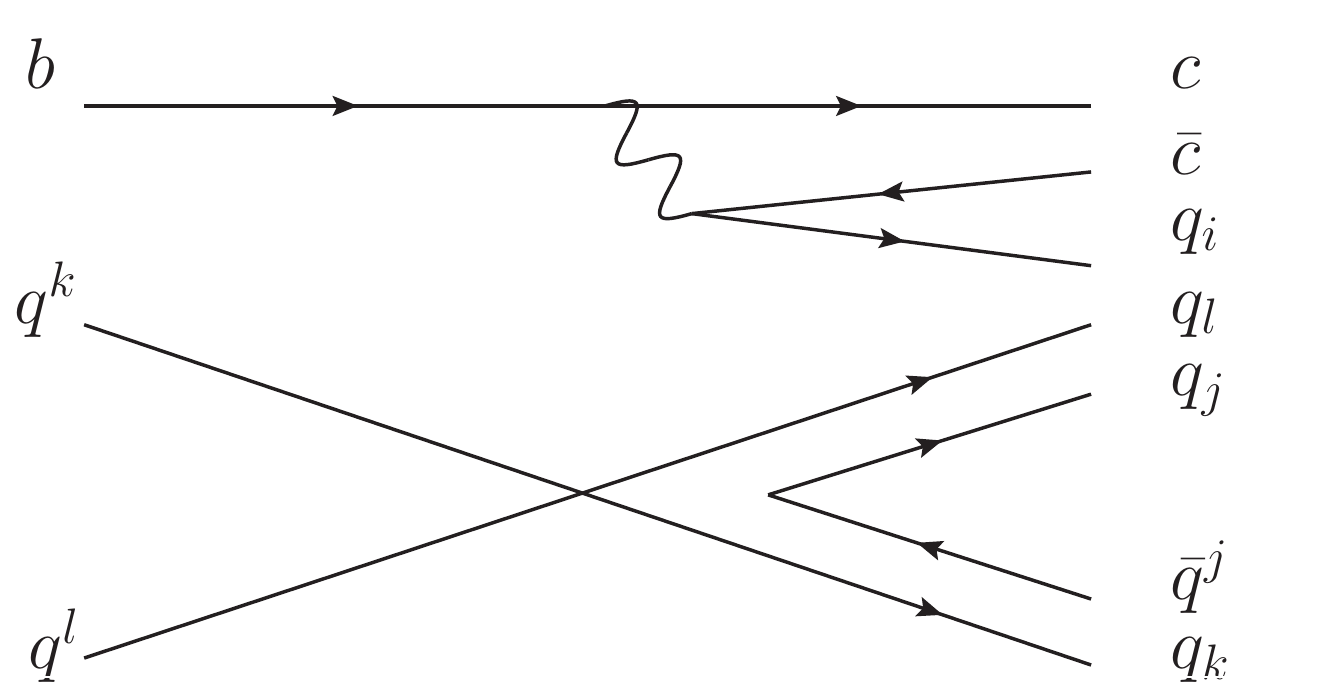}
}
%\centering{
%     \includegraphics[width=0.5\textwidth,natwidth=610,natheight=642]{BbtoPiP.pdf}
%}
\caption{External $W$-emission (a) and internal $W$-emission diagrams (b) and (c) for the bottom baryon decays to a pseudoscalar meson and a pentaquark. Note that the amplitude of Fig. (c) is the same as that of Fig. (b) up to a sign, since the former corresponds to switching $k$ and $l$ in the initial diquark state of the latter, which is antisymmetric (symmetric) under the interchange of two light quarks of the antitriplet (sextet) bottom baryon.  }
\label{fig:BbtoPiP}
\end{figure}
%%%%%%%%%%%%%%%%%%%%%%%%%%%%%%%%%%%%%%%%%%%%%%%%

The new charmonium-like pentaquarks were observed by LHCb as $J/\psi+p$ resonances produced in $\Lambda_b^0$ decays.  Since a proton transforms as an octet under SU(3), the pentaquarks should also belong to octet multiplets.
For octet pseudoscalar and pentaquark multiplets, we write%~\cite{text}
\be
\Pi&=& \left(
\begin{array}{ccc}
{{\pi^0}\over\sqrt2}+{{\eta_8}\over\sqrt6}&{\pi^+} &{K^+}\\
{\pi^-}&-{{\pi^0}\over\sqrt2}+{{\eta_8}\over\sqrt6}&{K^0}\\
{K^-}  &{\overline K {}^0}&-\sqrt{2\over3}{\eta_8}
\end{array}
\right),
\non\\
{\mathcal P}&=& \left(
\begin{array}{ccc}
{P_{\Sigma^0}\over\sqrt2}+{P_{\Lambda}\over\sqrt6}
       &P_{\Sigma^+}
       &P_{p}
       \\
P_{\Sigma^-}
       &-{P_{\Sigma^0}\over\sqrt2}+{P_{\Lambda}\over\sqrt6}
       &{P_n}
       \\
P_{\Xi^-}
       &P_{\Xi^0}
       &-\sqrt{2\over3}{P_\Lambda}
\end{array}
\right), \label{eq:octet}
\en
where we have denoted $P_p$ as the hidden-charm pentaquark with the same light quark content as the proton and likewise for other pentaquark fields.
Note that the ${\cal P}^j_k$ has the flavor structure $\bar c c( q^j q^a
q^b \epsilon_{abk}-\frac{1}{3}\,\delta^j_k \epsilon^{abl} q^l q^a
q^b)$.
To match the flavor of $q_i \bar q^j$ and $\bar c c q_j q_k q_l$ final states in bottom baryon decays as depicted in Fig.~\ref{fig:BbtoPiP}(a), for example,
we use~\footnote{The procedure is similar to the one used in \cite{Chua:2003it, Chua:2013zga}.}
\be
q_i\bar q^j\to\Pi^j_i,
\quad
\bar c c q_j q_k q_l\to \epsilon_{jka} \overline {\cal P}^a_l,\,\
                 \epsilon_{jal} \overline {\cal P}^a_k,\,\
                 \epsilon_{akl} \overline {\cal P}^a_j,
\label{eq:qqq}
\en
as the corresponding rules in obtaining $H_{\rm eff}$.~\footnote{We use subscript and superscript according
to the field convention. For example, we assign a superscript
(subscript) to the initial (final) quark state $q^k$~($q_k$).}
Note that the right-hand-side of the above equation contains all possible permutations of $j,k,l$.
In fact, not all terms are
independent as there is an identity,
$
\epsilon_{ika} \overline {\cal P}^a_l+\epsilon_{ial} \overline
{\cal P}^a_k+\epsilon_{akl} \overline {\cal
P}^a_i=0.
$
Therefore, for the $\bar c c q_k q_j q_l$
configuration we only need two independent terms.
We will choose two of them for our convenience.

We now come to the initial state. The low-lying bottom baryons can be classified into an antitriplet ${\cal B}_a$ and a sextet ${\cal B}^{kl}$ under SU(3):
\be
{\cal B}_a=(\Lambda_b^0, \Xi_b^0, \Xi_b^-), \qquad \quad {\cal B}^{kl}&=& \left(
\begin{array}{ccc}
 \Sigma_b^+ & {\Sigma_b^0\over\sqrt2} & {\Xi_b^{\prime 0}\over \sqrt2} \\
{\Sigma_b^0\over\sqrt2} & \Sigma_b^- & {\Xi_b^{\prime-}\over \sqrt2} \\
{\Xi_b^{\prime0}\over \sqrt2} & {\Xi_b^{\prime-}\over \sqrt2} & \Omega_b^-
\end{array}
\right).
\en
While all the bottom baryons in the ${\bf \bar 3}$ representation
decay weakly,  only $\Omega_b^-$ in the {\bf 6} representation decays weakly.
We can project $b q^k q^l$ in Fig.\ref{fig:BbtoPiP}  to an antitriplet state or a sextet state according to the following rule
\be
b q^k q^l\to \epsilon^{kla}{\cal B}_a,\quad {\cal B}^{kl}.
\en

We can now write down the effective Hamiltonian in terms of hadronic degrees of freedom. Using the above corresponding rules, we obtain~\footnote{To incorporate the SU(3)-singlet state $\eta_1$, we will make use of $U(3)$ symmetry by adding $\delta^j_i\eta_1/\sqrt3$ to the $\Pi^j_i$ matrix elements.}
\be
H_{\rm eff}&=&\epsilon^{kla}{\cal B}_a H^i \Pi^j_i
(\epsilon_{jkb}\bar{\cal P}^b_l T_{1'}+\epsilon_{jbl}\bar{\cal P}^b_k T_{1''})
\non\\
&&+\epsilon^{kla}{\cal B}_a H^i \Pi^j_l
(\epsilon_{ikb}\bar{\cal P}^b_j T_{2}+\epsilon_{bkj}\bar{\cal P}^b_i T_3)
\non\\
&&+{\cal B}^{kl} H^i \Pi^j_i
(\epsilon_{jka}\bar{\cal P}^a_l t_{1'}+\epsilon_{jal}\bar{\cal P}^a_k t_{1''})
\non\\
&&+{\cal B}^{kl} H^i \Pi^j_l
(\epsilon_{ika}\bar{\cal P}^a_j t_{2}+\epsilon_{akj}\bar{\cal P}^a_i t_3),
\en
where $T_{1',1''}(t_{1',1''})$ terms correspond to the external $W$-emission of Fig.~1(a), and $T_{2,3}(t_{2,3})$ to the internal $W$-emission of Fig.~1(b).
By interchanging $k$ and $l$ indices in the second and fourth terms
%of the terms in the parenthesis of the $T_2(t_2)$ and $T_3(t_3)$ terms
and replacing $T_{2,3}(t_{2,3})$ by $T_{4,5}(t_{4,5})$, we obtain the  contributions from Fig. 1(c). However, the additional $T_{4,5}(t_{4,5})$ terms are identical to the $T_{2,3}(t_{2,3})$ ones up to a sign and can be absorbed in $T_{2,3}(t_{2,3})$.
With the help of the spurion field $H^i$, this effective Hamiltonian has the same SU(3) property as the one in terms of quark fields.
Defining $T_1\equiv -T_{1'}-T_{1''}$ and $t_1\equiv t_{1'}-t_{1''}$, we can recast the above Hamiltonian as
\be
H_{\rm eff}&=&{\cal B}_a H^i \Pi^j_i\bar{\cal P}^a_j (T_1-T_2)
+{\cal B}_a H^a \Pi^j_i \bar{\cal P}^i_j T_2
+{\cal B}_a H^i \Pi^j_j \bar{\cal P}^a_i T_3
-{\cal B}_a H^i \Pi^a_l \bar{\cal P}^l_i T_3
\non\\
&&
+{\cal B}^{kl} H^i \Pi^j_i\epsilon_{jka}\bar{\cal P}^a_l t_1
+{\cal B}^{kl} H^i \Pi^j_l\epsilon_{ika}\bar{\cal P}^a_j t_2
+{\cal B}^{kl} H^i \Pi^j_l\epsilon_{akj}\bar{\cal P}^a_i t_3.
\en
%It is true that this is the most generic terms that we can almost directly write down from the symmetry argument and some of the above steps can be bypassed. However, the above procedure allow one to see clearly how these terms correspond to the flavor flow diagram Fig.~\ref{fig:BbtoPiP} and the same procedure can be used when further terms, such as penguin contributions, are included in future studies.

%%%%%%%%%%%%%%%%%%%%%%%%%%%%%%%%%%%%%%%%%%%%%%%%%%%%%%%%%%%
{\footnotesize
\begin{table}[t!]
\caption{\label{tab:amplitudes1}
The decay amplitudes of bottom baryon decays to a pseudoscalar and a octet pentaquark. $T_i$ and $t_i$ $(i=1,2,3)$ are $\Delta S=-1$ transition amplitudes for antitriplet and sextet bottom baryons, respectively. }
\begin{ruledtabular}
\begin{tabular}{lc lc}
Process
          & Amplitude
          & Process
          & Amplitude
          \\
\hline
$\Lambda_b^0\to P_p K^-$
          & $T_1$
          & $\Lambda_b^0\to P_{n} \bar K^0$
          & $T_1$
          \\
$\Lambda_b^0\to P_{\Lambda}\eta$
          & $\frac{1}{3}[(2T_1+T_2-2 T_3)\cos\theta$
          & $\Lambda_b^0\to P_{\Lambda}\eta'$
          & $\frac{1}{3}[-\sqrt2(T_1-T_2+2 T_3)\cos\theta$
          \\
%$\Lambda_b^0\to P_{\Lambda}\eta'$
          & $+(2 T_1+T_2-2T_3)\sin\theta\big]$
          & %$\Lambda_b^0\to P_{\Lambda}\eta$
          & $+\sqrt2(T_1-T_2+2T_3)\sin\theta\big]$
          \\
$\Lambda_b^0\to P_{\Sigma^+}\pi^-$
          & $T_2$
          & $\Lambda_b^0\to P_{\Sigma^-}\pi^+$
          & $T_2$
          \\
$\Lambda_b^0\to P_{\Xi^0} K^0$
          & $T_2-T_3$
          & $\Lambda_b^0\to P_{\Xi^-} K^+$
          & $T_2-T_3$
          \\
$\Lambda_b^0\to P_{\Sigma^0}\pi^0$
          &  $T_2$
          & $\Lambda_b^0\to P_{\Lambda}\pi^0$
          & 0
          \\
$\Lambda_b^0\to P_{\Sigma^0}\eta$
          & 0
          & $\Lambda_b^0\to P_{\Sigma^0}\eta'$
          & 0
          \\
$\Xi^0_b\to P_{\Sigma^+}K^-$
          & $T_1-T_2$
          & $\Xi^0_b\to P_{\Sigma^0}\bar K^0$
          & $\frac{1}{\sqrt2}(-T_1+T_2)$
          \\
$\Xi^0_b\to P_{\Xi^0}\eta$
          & $-\frac{1}{\sqrt6}(2 T_1-2 T_2+T_3)\cos\theta$
          & $\Xi^0_b\to P_{\Xi^0}\eta'$
          & $\frac{1}{\sqrt3}(T_1-T_2+2T_3)\cos\theta$
          \\
%$\Xi^0_b\to P_{\Xi^0}\eta'$
          & $-\frac{1}{\sqrt6}(2T_1-2T_2+ T_3)\sin\theta$
          & %$\Xi^0_b\to P_{\Xi^0}\eta$
          & $-\frac{1}{\sqrt3}(T_1-T_2+2 T_3)\sin\theta$
          \\
$\Xi_b^0\to P_{\Xi^-}\pi^+$
          & $-T_3$
          & $\Xi_b^0\to P_{\Xi^0}\pi^0$
          &  $\frac{1}{\sqrt2} T_3$
          \\
$\Xi_b^0\to P_{\Lambda}\bar K^0$
          & ${1\over \sqrt6}(T_1-T_2+2 T_3)$
          \\
$\Xi^-_b\to P_{\Sigma^-} \bar K^0$
          & $T_1-T_2$
          & $\Xi^-_b\to P_{\Sigma^0} K^-$
          & $\frac{1}{\sqrt2} (T_1-T_2)$
          \\
$\Xi_b^-\to P_{\Xi^-}\pi^0$
          & $-\frac{1}{\sqrt2} T_3$
          & $\Xi_b^-\to P_{\Xi^0}\pi^-$
          & $-T_3$
          \\
$\Xi^-_b\to P_{\Xi^-}\eta$
          & $-\frac{1}{\sqrt6} (2T_1-2T_2+T_3)\cos\theta$
          & $\Xi^-_b\to P_{\Xi^-}\eta'$
          & $\frac{1}{\sqrt3} (T_1-T_2+2T_3)\cos\theta$
          \\
          & $-\frac{1}{\sqrt3} (T_1-T_2+2T_3)\sin\theta$
          & %$\Xi^-_b\to P_{\Xi^-}\eta$
          & $-\frac{1}{\sqrt6} (2T_1-2T_2+T_3)\sin\theta$
          \\
$\Xi^-_b\to P_{\Lambda} K^-$
          & $\frac{1}{\sqrt6} (T_1-T_2+2T_3)$
          \\
$\Omega_b^-\to P_{\Xi^-} \bar K^0$
          & $t_1-t_3$
          & $\Omega_b^-\to P_{\Xi^0} K^-$
          & $-t_1+t_3$
           \\
%$\Omega_b^-\to P_{\Xi^{*-}} \bar K^0$
 %         & $\frac{1}{\sqrt3} \tilde T(6)$
 %         & $\Omega_b^-\to P_{\Xi^{*0}} K^-$
 %         & $\frac{1}{\sqrt3} \tilde T(6)$
 %         \\
%$\Omega_b^-\to P_{\Omega^{-}} \eta$
%          & $-\frac{1}{\sqrt3}\left(\sqrt2\cos\theta+\sin\theta\right) \tilde T(6)$
%          & $\Omega_b^-\to P_{\Omega^{-}} \eta'$
%          & $\frac{1}{\sqrt3}\left(\cos\theta-\sqrt2\sin\theta\right) \tilde T(6)$
\end{tabular}
\end{ruledtabular}
\end{table}
}
%%%%%%%%%%%%%%%%%%%%%%%%%%%%%%%%%%%%%%%%%%%%%%%%%%%%%%%%%%%%%%

%%%%%%%%%%%%%%%%%%%%%%%%%%%%%%%%%%%%%%%%%%%%%%%%%%%%%%%%%%%
{\footnotesize
\begin{table}[t!]
\caption{\label{tab:amplitudes2}
The decay amplitudes of bottom baryon decays to a pseudoscalar and a octet pentaquark. $T'_i$ and $t'_i$ $(i=1,2,3)$ are $\Delta S=0$ transition amplitudes for antitriplet and sextet bottom baryons, respectively.}
\begin{ruledtabular}
\begin{tabular}{lc lc}
Process
          & Amplitude
          & Process
          & Amplitude
          \\
\hline
$\Lambda_b^0\to P_p\pi^-$
          & $T'_1-T'_2$
          & $\Lambda_b^0\to P_n \pi^0$
          & $-\frac{1}{\sqrt2} (T'_1-T'_2)$
          \\
$\Lambda_b^0\to P_{\Sigma^0}K^0$
          & $\frac{1}{\sqrt2} T'_3$
          & $\Lambda_b^0\to P_{\Sigma^-}K^+$
          & $-T'_3$
          \\
$\Lambda_b^0\to P_n\eta$
          & $\left(\frac{\cos\theta}{\sqrt6}-\frac{\sin\theta}{\sqrt3}\right)(T'_1-T'_2+2T'_3)$
          & $\Lambda_b^0\to P_n\eta'$
          & $\left(\frac{\cos\theta}{\sqrt3}+\frac{\sin\theta}{\sqrt6}\right)(T'_1-T'_2+2T'_3)$
          \\
$\Lambda_b^0\to P_{\Lambda} K^0$
          & $-\frac{1}{\sqrt6} (2T'_1-2T'_2+T'_3)$
          \\
$\Xi^0_b\to P_{\Sigma^+}\pi^-$
          & $T'_1$
          & $\Xi_b^0\to P_{\Sigma^0}\pi^0$
          & $\frac{1}{2}(T'_1+T'_2-T'_3)$
          \\
$\Xi^0_b\to P_{\Xi^0}K^0$
          & $T_1'$
          & $\Xi^0_b\to P_{\Xi^-}K^+$
          & $T'_2$
          \\
$\Xi^0_b\to P_{\Lambda}\eta$
          & $\frac{1}{6} \cos\theta (T'_1+5T'_2-T'_3)$
          & $\Xi^0_b\to P_{\Lambda}\eta'$
          & $\frac{1}{3\sqrt2} \cos\theta (T'_1-T'_2+2T'_3)$
          \\
%$\Xi^0_b\to P_{\Lambda}\eta$
          &  $-\frac{1}{3\sqrt2} \sin\theta (T'_1-T'_2+2T'_3)$
          & %$\Xi^0_b\to P_{\Lambda}\eta'$
          & $+\frac{1}{6} \sin\theta (T'_1+5T'_2-T'_3)$
          \\
$\Xi^0_b\to P_{\Sigma^0}\eta$
          & $\frac{1}{2\sqrt3} \cos\theta (-T'_1+T'_2+T'_3)$
          & $\Xi^0_b\to P_{\Sigma^0}\eta'$
          & $-\frac{1}{\sqrt6} \cos\theta (T'_1-T'_2+2T'_3)$
          \\
%$\Xi^0_b\to P_{\Sigma^0}\eta$
          & $+\frac{1}{\sqrt6} \sin\theta (T'_1-T'_2+2T'_3)$
          & %$\Xi^0_b\to P_{\Sigma^0}\eta'$
          & $\frac{1}{2\sqrt3} \sin\theta (-T'_1+T'_2+T'_3)$
          \\
$\Xi^0_b\to P_{p}K^-$
          & $T'_2$
          & $\Xi^0_b\to P_{n}\bar K^0$
          & $T'_2-T'_3$
          \\
$\Xi^0_b\to P_{\Sigma^-}\pi^+$
          & $T'_2-T'_3$
          & $\Xi^0_b\to P_{\Lambda}\pi^0$
          & $\frac{1}{2\sqrt3} (-T'_1+T'_2+T'_3)$
          \\
$\Xi^-_b\to P_{\Xi^-} K^0$
          & $T'_1-T'_2$
          & $\Xi_b^-\to P_{n}K^-$
          & $-T'_3$
          \\
$\Xi^-_b\to P_{\Sigma^-} \eta$
          & $\frac{1}{\sqrt6}\cos\theta(T'_1-T'_2-T'_3)$
          & $\Xi^-_b\to P_{\Sigma^-} \eta'$
          & $\frac{1}{\sqrt3}\cos\theta(T'_1-T'_2+2T'_3)$
          \\
%$\Xi^-_b\to P_{\Sigma^-} \eta$
          & $-\frac{1}{\sqrt3}\sin\theta(T'_1-T'_2+2T'_3)$
          & %$\Xi^-_b\to P_{\Sigma^-} \eta'$
          & $+\frac{1}{\sqrt6}\sin\theta(T'_1-T'_2-T'_3)$
          \\
$\Xi^-_b\to P_{\Sigma^-} \pi^0$
          & -$\frac{1}{\sqrt2} (T'_1-T'_2+T'_3)$
          & $\Xi^-_b\to P_{\Sigma^0} \pi^-$
          & $\frac{1}{\sqrt2} (T'_1-T'_2+T'_3)$
          \\
$\Xi^-_b\to P_{\Lambda} \pi^-$
          & $\frac{1}{\sqrt6} (T'_1-T'_2-T'_3)$
          \\
$\Omega_b^-\to P_{\Xi^-} \pi^0$
          & $-\frac{1}{\sqrt2}t'_1$
          & $\Omega_b^-\to P_{\Xi^0} \pi^-$
          &   $-t'_1$
          \\
$\Omega_b^-\to P_{\Xi^-} \eta$
          & $\frac{1}{\sqrt6}\cos\theta (t'_1-2t'_2)$
          & $\Omega_b^-\to P_{\Xi^-} \eta'$
          & $\frac{1}{\sqrt3}\cos\theta(t'_1+t'_2)$
          \\
%$\Omega_b^-\to P_{\Xi^-} \eta$
          & $-\frac{1}{\sqrt3}\sin\theta (t'_1+t'_2)$
          & %$\Omega_b^-\to P_{\Xi^-} \eta'$
          & $+\frac{1}{\sqrt6}\sin\theta (t'_1-2t'_2)$
          \\
$\Omega_b^-\to P_{\Sigma^-}\bar K^0$
          & $t'_2-t'_3$
          & $\Omega_b^-\to P_{\Sigma^0}K^-$
          & $\frac{1}{\sqrt2}(t'_2-t'_3)$
          \\
$\Omega_b^-\to P_{\Lambda}K^-$
          & $\frac{1}{\sqrt6}(t'_2+t'_3)$
           \\
\end{tabular}
\end{ruledtabular}
\end{table}
}
%%%%%%%%%%%%%%%%%%%%%%%%%%%%%%%%%%%%%%%%%%%%%%%%%%%%%%%%%%%%%%

%%
%%
%%
%%
%%%%%%%%%%%%%%%%%%%%%%%%%%%%%%%%%%%%%%%%%%%%%%%%%%%%%%%%%%%
{\footnotesize
\begin{table}[t!]
\caption{\label{tab:amplitudes3}
The decay amplitudes of bottom baryon decays to a pseudoscalar and a decuplet pentaquark. $\tilde T_2$ and $\tilde t_i$ $(i=1,2)$ are $\Delta S=-1$ transition amplitudes for antitriplet and sextet bottom baryons, respectively.}
\begin{ruledtabular}
\begin{tabular}{lc lc}
Process
          & Amplitude
          & Process
          & Amplitude
          \\
\hline
$\Lambda_b^0\to P_{\Xi^{*0}} K^0$
          & $\frac{1}{\sqrt3}\tilde T_2$
          &$\Lambda_b^0\to P_{\Xi^{*-}}K^+$
          & $-\frac{1}{\sqrt3}\tilde T_2$
          \\
 $\Lambda_b^0\to P_{\Sigma^{*0}} \pi^0$
          & $-\frac{1}{\sqrt3}\tilde T_2$
          & $\Lambda_b^0\to P_{\Sigma^{*+}} \pi^-$
          & $\frac{1}{\sqrt3}\tilde T_2$
          \\
$\Lambda_b^0\to P_{\Sigma^{*-}}\pi^+$
          & $-\frac{1}{\sqrt3}\tilde T_2$
          &
          &
          \\
$\Xi^0_b\to P_{\Sigma^{*0}}\bar K^0$
          & $\frac{1}{\sqrt6}\tilde T_2$
          & $\Xi^0_b\to P_{\Sigma^{*+}}K^-$
          & $-\frac{1}{\sqrt3}\tilde T_2$
          \\
$\Xi^0_b\to P_{\Xi^{*0}}\eta$
          & $\frac{1}{\sqrt2}\cos\theta\tilde T_2$
          & $\Xi^0_b\to P_{\Xi^{*0}}\eta'$
          & $\frac{1}{\sqrt2}\sin\theta\tilde T_2$
          \\
$\Xi^0_b\to P_{\Xi^{*0}}\pi^0$
          & $\frac{1}{\sqrt6}\tilde T_2$
          & $\Xi^0_b\to P_{\Xi^{*-}}\pi^+$
          & $\frac{1}{\sqrt3}\tilde T_2$
          \\
$\Xi_b^0\to P_{\Omega^-}K^+$
          & $\tilde T_2$
          &
          &
          \\
$\Xi^-_b\to P_{\Sigma^{*-}} \bar K^0$
          & $\frac{1}{\sqrt3}\tilde T_2$
          & $\Xi^-_b\to P_{\Sigma^{*0}} K^-$
          & $\frac{1}{\sqrt6}\tilde T_2$
          \\
$\Xi^-_b\to P_{\Xi^{*-}}\eta$
          & $-\frac{1}{\sqrt2}\cos\theta\tilde T_2$
          & $\Xi^-_b\to P_{\Xi^-}\eta'$
          & $-\frac{1}{\sqrt2}\sin\theta\tilde T_2$
          \\
$\Xi^-_b\to P_{\Xi{*-}} \pi^0$
          & $\frac{1}{\sqrt6}\tilde T_2$
          & $\Xi^-_b\to P_{\Xi^{*0}} \pi^-$
          & $-\frac{1}{\sqrt3}\tilde T_2$
          \\
$\Xi_b^-\to P_{\Omega^-}K^0$
          & $-\tilde{T}_2$
          \\
$\Omega_b^-\to P_{\Xi^{*-}} \bar K^0$
          & $\frac{1}{\sqrt3} (\tilde t_1+\tilde t_2)$
          & $\Omega_b^-\to P_{\Xi^{*0}} K^-$
          & $\frac{1}{\sqrt3}(\tilde t_1+\tilde t_2)$
          \\
$\Omega_b^-\to P_{\Omega^{-}} \eta$
          & $-\frac{1}{\sqrt3}\left(\sqrt2\cos\theta+\sin\theta\right) (\tilde t_1+\tilde t_2)$
          & $\Omega_b^-\to P_{\Omega^{-}} \eta'$
          & $\frac{1}{\sqrt3}\left(\cos\theta-\sqrt2\sin\theta\right) (\tilde t_1+\tilde t_2)$
\end{tabular}
\end{ruledtabular}
\end{table}
}
%%%%%%%%%%%%%%%%%%%%%%%%%%%%%%%%%%%%%%%%%%%%%%%%%%%%%%%%%%%%%%

%%%%%%%%%%%%%%%%%%%%%%%%%%%%%%%%%%%%%%%%%%%%%%%%%%%%%%%%%%%
{\footnotesize
\begin{table}[t!]
\caption{\label{tab:amplitudes4}
The decay amplitudes of bottom baryon decays to a pseudoscalar and a decuplet pentaquark. $\tilde T'_2$ and $\tilde t'_i$ $(i=1,2)$ are $\Delta S=0$ transition amplitudes for antitriplet and sextet bottom baryons, respectively.}
\begin{ruledtabular}
\begin{tabular}{lc lc}
Process
          & Amplitude
          & Process
          & Amplitude
          \\
\hline
$\Lambda_b^0\to P_{\Sigma^{*0}}K^0$
          & $\frac{1}{\sqrt6} \tilde T'_2$
          & $\Lambda_b^0\to P_{\Sigma^{*-}}K^+$
          & $-\frac{1}{\sqrt3} \tilde T'_2$
          \\
$\Lambda_b^0\to P_{\Delta^0} \pi^0$
          & $-\sqrt{\frac{2}{3}} \tilde T'_2$
          & $\Lambda_b^0\to P_{\Delta^-}\pi^+$
          & $-\tilde T'_2$
          \\
$\Lambda_b^0\to P_{\Delta^+}\pi^-$
          & $\frac{1}{\sqrt3}\tilde T'_2$
          &
          &
          \\
$\Xi^0_b\to P_{\Sigma^{*0}}\pi^0$
          & $\frac{1}{2\sqrt3}\tilde T'_2$
          & $\Xi_b^0\to P_{\Sigma^{*-}}\pi^+$
          & $\frac{1}{\sqrt3}\tilde T'_2$
          \\
$\Xi^0_b\to P_{\Sigma^{*0}}\eta$
          & $\frac{1}{2} \cos\theta \tilde T'_2$
          & $\Xi^0_b\to P_{\Sigma^{*0}}\eta'$
          & $\frac{1}{2} \sin\theta \tilde T'_2$
          \\
$\Xi^0_b\to P_{\Delta^0}\bar K^0$
          & $-\frac{1}{\sqrt3}\tilde T'_2$
          & $\Xi^0_b\to P_{\Delta^+}K^-$
          & $-\frac{1}{\sqrt3}\tilde T'_2$
          \\
$\Xi^0_b\to P_{\Xi^{*-}}\bar K^+$
          & $\frac{1}{\sqrt3}\tilde T'_2$
          &
          &
          \\
$\Xi^-_b\to P_{\Delta^-} \bar K^0$
          & $\tilde T'_2$
          & $\Xi^-_b\to P_{\Delta^0} K^-$
          & $\frac{1}{\sqrt3} \tilde T'_2$
          \\
$\Xi^-_b\to P_{\Sigma^-} \eta$
          & $-\frac{1}{\sqrt2}\cos\theta \tilde T'_2$
          & $\Xi^-_b\to P_{\Sigma^-} \eta'$
          & $-\frac{1}{\sqrt2}\sin\theta\tilde T'_2$
          \\
$\Xi^-_b\to P_{\Sigma^{*-}} \pi^0$
          & $\frac{1}{\sqrt6}\tilde T'_2$
          &$\Xi_b^-\to P_{\Sigma^{*0}}\pi^-$
          & $-\frac{1}{\sqrt6}\tilde T'_2$
          \\
$\Xi^-_b\to P_{\Xi^{*-}} K^0$
          & $-\frac{1}{\sqrt3}\tilde T'_2$
          &
          &
          \\
$\Omega_b^-\to P_{\Xi^{*-}} \eta$
          & $\frac{1}{3\sqrt2}\cos\theta (\tilde t'_1-2\tilde t'_2)$
          & $\Omega_b^-\to P_{\Xi^{*-}} \eta'$
          & $\frac{1}{3}\cos\theta(\tilde t'_1+\tilde t'_2)$
          \\
%$\Omega_b^-\to P_{\Xi^{*-}} \eta$
          & $-\frac{1}{3}\sin\theta (\tilde t'_1+\tilde t'_2)$
          & %$\Omega_b^-\to P_{\Xi^{*-}} \eta'$
          & $+\frac{1}{3\sqrt2}\sin\theta(\tilde t'_1-2\tilde t'_2)$
          \\
$\Omega_b^-\to P_{\Xi^{*-}} \pi^0$
          & $-\frac{1}{\sqrt6}\tilde t'_1$
          & $\Omega_b^-\to P_{\Xi^{*0}} \pi^-$
          & $\frac{1}{\sqrt3}\tilde t'_1$
          \\
$\Omega_b^-\to P_{\Sigma^0} K^-$
          & $\frac{1}{\sqrt6}\tilde t'_2$
          & $\Omega_b^-\to P_{\Sigma^{*-}} \bar K^0$
          & $\frac{1}{\sqrt3}\tilde t'_2$
           \\
$\Omega_b^-\to P_{\Omega^-} K^0$
          & $\tilde t'_1$
          \\
\end{tabular}
\end{ruledtabular}
\end{table}
}
%%%%%%%%%%%%%%%%%%%%%%%%%%%%%%%%%%%%%%%%%%%%%%%%%%%%%%%%%%%%%%

We may further consider the decuplet pentaquark field ${\cal P}^{ijk}$
with
%$\bar q^l \bar q^j \bar q^m$ flavor
%is created by a ${\cal D}^{jlm}$ field, where
%${\cal D}^{jlm}$ is
%the familiar decuplet field with
${\cal P}^{111}=P_{\Delta^{++}}$,
${\cal P}^{112}=P_{\Delta^{+}}/\sqrt3$,
${\cal P}^{122}=P_{\Delta^0}/\sq3$,
${\cal P}^{222}=P_{\Delta^-}$,
${\cal P}^{113}=P_{\Sigma^{*+}}/\sq3$,
${\cal P}^{123}=P_{\Sigma^{*0}}/\sq6$,
${\cal P}^{223}=P_{\Sigma^{*-}}/\sq3$,
${\cal P}^{133}=P_{\Xi^{*0}}/\sq3$,
${\cal P}^{233}=P_{\Xi^{*-}}/\sq3$
and ${\cal P}^{333}=P_{\Omega^-}$.
%(see, for example~\cite{text}).
The corresponding rule is
\be
\bar c c q_j q_k q_l\to \bar{\cal P}_{jkl},
\en
and, consequently, the related weak Hamiltonian is
\be
H_{\rm eff}&=&\epsilon^{kla}{\cal B}_a H^i\Pi^j_l\bar{\cal P}_{ijk}\tilde T_2
+{\cal B}^{kl} H^i \Pi^j_i\bar{\cal P}_{jkl} \tilde t_1
+{\cal B}^{kl} H^i \Pi^j_l\bar{\cal P}_{ijk} \tilde t_2,
\en
where $\tilde t_1$ and $\tilde T_2(\tilde t_2)$ correspond to Fig. 1(a) and 1(b) [including 1(c)], respectively. It should be stressed that an anti-triplet bottom baryon can decay to a decuplet pentaquark only through Fig. 1(b) [with 1(c) as well], since the light quark flavor is antisymmetric in the initial state, while it is symmetric in the final state. As a result, there is no contribution from Fig. 1(a).

Decay amplitudes of bottom baryon decays to a pseudoscalar meson and a pentaquark in the antitriplet and sextet are listed in Tables \ref{tab:amplitudes1}-\ref{tab:amplitudes2} and Tables \ref{tab:amplitudes3}-\ref{tab:amplitudes4}, respectively, for $\Delta S=-1\,(0)$ transitions.
%Note that we only consider bottomed baryons that decay mainly through weak interaction, where these bottomed baryons may live long enough to undergo the pseudoscalar meson and pentaquark decays.
%
The SU(3) octet and singlet states $\eta_8$ and $\eta_1$, respectively, are related to the physical $\eta$ and $\eta'$ states via
\be
 \left(\matrix{ \eta_8 \cr \eta_1\cr}\right)=\left(\matrix{ \cos\theta & \sin\theta \cr
 -\sin\theta & \cos\theta\cr}\right)\left(\matrix{\eta \cr \eta'
 \cr}\right).
\en
The most recent experimental determination of the $\eta\!-\!\eta'$ mixing angle is $\theta=-(14.3\pm 0.6)^\circ$ from KLOE \cite{KLOE},
which is indeed close to the original theoretical and phenomenological estimates of $-12.5^\circ$ and $(-15.4\pm1.0)^\circ$, respectively, made by Feldmann, Kroll and Stech \cite{FKS}.
It is interesting to notice that the decay amplitudes of $\Lambda_b^0\to P_{\Lambda}\pi^0,~P_{\Sigma^0}\eta$ and $P_{\Sigma^0}\eta'$ vanish in the  SU(3) limit (see Table \ref{tab:amplitudes1}). Many relations can be read off from Tables \ref{tab:amplitudes1}-\ref{tab:amplitudes4}, for example,
\be \label{eq:relation}
&& A(\Lambda_b^0\to P_{\Sigma^0}\pi^0)=A(\Lambda_b^0\to P_{\Sigma^+}\pi^-)=A(\Lambda_b^0\to P_{\Sigma^-}\pi^+), \non \\
&& A(\Xi_b^0\to P_{\Sigma^+}K^-)-\sqrt2 A(\Xi_b^0\to P_{\Sigma^0}\bar K^0)=A(\Xi_b^-\to P_{\Sigma^-}\bar K^0)=\sqrt2 A(\Xi_b^-\to P_{\Sigma^0}K^-), \non \\
&& A(\Xi_b^0\to P_{\Xi^-}\pi^+)=-\sqrt2 A(\Xi_b^0\to P_{\Xi^0}\pi^0)=\sqrt2 A(\Xi_b^-\to P_{\Xi^-}\pi^0)=A(\Xi_b^-\to P_{\Xi^0}\pi^-), \non \\
&& A(\Xi_b^0\to P_{\Sigma^+}\pi^-)=A(\Xi_b^0\to P_{\Xi^0}K^0),\quad A(\Xi_b^0\to P_{\Xi^-}K^+)=A(\Xi_b^0\to P_{p}K^-), \\
&& \sqrt2 A(\Xi_b^0\to P_{\Sigma^0}\eta)=-A(\Xi_b^-\to P_{\Sigma^-}\eta), \quad \sqrt2 A(\Xi_b^0\to P_{\Sigma^0}\eta')=-A(\Xi_b^-\to P_{\Sigma^-}\eta'). \non
\en
There are totally 19 decay channels for the antitriplet bottom baryon decays to a pseudoscalar and a decuplet pentaquark as listed in Tables \ref{tab:amplitudes3} and \ref{tab:amplitudes4}. Their decay amplitudes are governed by $\tilde T_2$ and $\tilde T'_2$ for $\Delta S=-1$ and $\Delta S$ transitions, respectively. Consequently, the decays of $(\Lambda_b^0,\Xi_b^0,\Xi_b^-)\to P_{\bf 10}+M$ are all related to each other.
The amplitudes of $\Omega_b^-\to P_{\bf 10}+M$ are proportional to $\tilde t_1+\tilde t_2$ and $\tilde t'_{1,2}$, respectively, for $\Delta S=-1$ and $\Delta S=0$ transitions. Hence, $\Omega_b^-$ is allowed to decay to a decuplet pentaquark through the external $W$-emission diagram.

Based on SU(3) flavor symmetry, weak decays of bottom baryons to a light pseudoscalar and an octet or decuplet pentaquark were also studied in \cite{Li:2015gta}. Several $U$-spin relations which relate $\Delta S=-1$ and $\Delta S=0$ amplitudes were derived there. In the work of \cite{Li:2015gta}, $(\Lambda_b^0,\Xi_b^0,\Xi_b^-)\to P_{\bf 8}+M$ decays are expressed in terms of eight unknown invariant amplitudes, while in our work they are expressed in terms of three invariant amplitudes $T_1$ and $T_{2,3}$ corresponding to the external $W$-emission and internal $W$-emission diagrams, respectively. Therefore, the physical pictures of invariant amplitudes are more transparent in our study. Nevertheless, as far as the relations between various modes are concerned such as Eq. (\ref{eq:relation}) in this work and Eqs.~(19), (22) and (23) in \cite{Li:2015gta}, we are in agreement with each other.

\section{Discussions}

In the absence of a dynamical model we are not able to estimate the absolute rate of the bottom baryon decays to a light pseudoscalar and a pentaquark. Nevertheless, under a plausible assumption on the relative importance of the external and internal $W$-emission diagrams, we can make a crude estimate on $\Gamma({\cal B}_b\to P_{\cal B}M)$ relative to $\Gamma(\Lambda_b^0\to P_p^+K^-)$.

The decay $\Lambda_b\to P_p K^-$ observed by LHCb receives contributions only from the external $W$-emission diagram Fig. 1(a).
Indeed, this contribution should be the dominating one in bottom baryon decays to a pseudoscalar and a pentaquark, since in internal $W$-emission diagrams [Figs. 1(b) and 1(c)], the three quarks ($c\,\bar c\,q_i$) produced directly from the $b$ quark decay are too energetic to form a pentaquark. As a consequence, it is likely that the internal $W$-emission diagram is suppressed relative to the external $W$-emission one. Under this hypothesis, we are going to show in Table~\ref{tab:rateratio} the estimation of rate ratios by assuming the dominant contributions from Fig. 1(a) and neglecting other contributions. It is true that all modes in Table I-IV should be searched, but the estimate on rates of some modes will also be useful at this moment.
The contributions of the neglected sub-leading terms can be studied later when more modes are discovered and detected.

%internal $W$-emission diagrams [Figs. 1(b) and 1(c)] require the pentaquark to bound the three quark antiquarks ($c \bar c q_i$) directly produced from the $b$ quark decay and they are, in general, too energetic to be bounded in the pentaquark.Indeed, it is lucky that the $\Lambda_b\to P_p K^-$ decay consists of Fig. 1(a) contribution.If it is the other way around, the decay may be too weak to be detected and it would not be the first detected mode.

In a two-body decay system, the decay rate and the center of mass momentum has the simple relation:
\be
\Gamma\propto |p_{cm}| |A|^2\propto |p_{cm}|^{2L+1},
\en
where $L$ is the orbital angular momentum quantum number of the two final-state particles. From the conservation of the angular momentum in the two-body decay, we have $1/2=|S-L|$ or, equivalently, $L=S\pm 1/2$, where $S$ is the spin of the pentaquark. For $S=1/2$, $L$ can only be $1$ or $2$, while for $S=5/2$, we have $L=2,\,3$. Since the best fit solution to the LHCb data yields $J^P=(3/2^-,5/2^+)$ for $P_p(4380)^+$ and $P_p(4450)^+$, respectively, we see that parity in the decay $\Lambda_b^0\to P_p(4380)^+ K^-$ is violated (conserved)  for $L=1\,(2)$ . Likewise, parity in the decay $\Lambda_b^0\to P_p(4450)^+ K^-$ is violated (conserved)  for $L=2\,(3)$. Since parity is not conserved in weak interactions, we, therefore, assign $L=1\,(2)$ to the $S=3/2\,(5/2)$ case.

%Note that some of these weak decays are parity conserving processes, while others are parity violating. Since the center of mass momenta of these processes are small, processes with smaller $L$ dominates. We, therefore, assign $L=1 (2)$ for the $S=3/2(5/2)$ case.

As for the pentaquark masses, we shall assume the same SU(3) breaking effects in the pentaquark sector and the low-lying baryon sector:
\be \label{eq:massrel}
m_{P_{{\cal B}'}}\simeq m_{P_{\cal B}}+m_{{\cal B}'}-m_{\cal B}.
\en
Moreover, we will assume that there are two different types of octet pentaquark multiplets with $J^P=3/2^-$ and $J^P=5/2^+$. The measured masses of $P_p(4380)^+$ and $P_p(4450)^+$ given in Eq. (\ref{eq:mass}) will be used as a benchmark to fix the mass of the other pentaquark $P_{\cal B}$ through Eq. (\ref{eq:massrel}).
The decay amplitudes of $\Delta S=-1,0$ processes are related to each other through the relation
\be
\left|\frac{T'}{T}\right|^2=\left|\frac{t'}{t}\right|^2=\left|\frac{\tilde t'}{\tilde t}\right|^2=\left|\frac{V_{cd}^*}{V_{cs}^*}\right|^2\simeq 0.053,
\label{eq: T'/T}
\en
where $V_{ij}$ are Cabibbo-Kobayashi-Maskawa matrix elements.
With the amplitudes given in Tables~I to IV
we are ready to estimate the rate ratios of some bottom baryon decays to a pseudoscalar and a pentaquark, namely, $\Gamma({\cal B}_b\to P_{\cal B}^{3/2(5/2)}M)/\Gamma(\Lambda_b^0\to P_p^{3/2(5/2)}K^-)$.
The results are shown in Table~\ref{tab:rateratio}.
We see that the $\Lambda_b\to P_p K^-$ rate has the largest rate among the pentaquark multiplet. It also has a very good detectability.
This may explain why it is the first mode observed by LHCb.
Some modes have rates of similar order to the $\Lambda_b\to P_p K^-$ one, for example, $\Xi_b^0\to P_{\Sigma^+}K^-$ and $\Xi_b^-\to P_{\Sigma^-}\bar K^0$, $\Omega_b^-\to P_{\Xi^-}\bar K^0$ and $\Omega_b^-\to P_{\Xi^0}K^-$.
These modes should be given the higher priority in experimental searches for pentaquarks.  Note that the $\Delta S=0$ modes are Cabibbo-suppressed (see Eq. (\ref{eq: T'/T})). A recent work in \cite{Hsiao:2015nna} suggested that the intrinsic charm content of the $\Lambda_b$ baryon may lead to a dominant mechanism for the pentaquark production in the decay $\Lambda_b^0\to P_p^+ K^-$. If this mechanism dominates, one will have the prediction \cite{Hsiao:2015nna}
\be
{\Br(\Lambda_b^0\to P_p^+\pi^-)\over \Br(\Lambda_b^0\to P_p^+K^-)}=0.8\pm0.1\,,
\en
to be compared with the value of $0.07\sim0.08$ in our model (see Table \ref{tab:rateratio}). Therefore, a measurement of $\Lambda_b^0\to P_p^+\pi^-$ will be very useful to discriminate among different models.

%%%%%%%%%%%%%%%%%%%%%%%%%%%%%%%%%%%%%%%%%%%%%%%%%%%%%%%%%%%
{\footnotesize
\begin{table}[t!]
\caption{\label{tab:rateratio}
Estimate of decay rate ratios of $\Gamma({\cal B}_b\to P_{\cal B}^{3/2(5/2)}M)/\Gamma(\Lambda_b^0\to P_p^{3/2(5/2)}K^-)$
for $\Delta S=-1$ (top) and $\Delta S=0$ (bottom) transitions based on the assumption that the external $W$-emission diagram Fig. 1(a) gives dominant contributions. Note that we have applied  Eq. (\ref{eq: T'/T}) for the $\Delta S=0$ case. }
\begin{ruledtabular}
\begin{tabular}{lclc}
Process
          & $\Gamma/\Gamma(\Lambda_b^0\to P^{3/2(5/2)}_p K^-)$
          & Process
          & $\Gamma/\Gamma(\Lambda_b^0\to P^{3/2(5/2)}_p K^-)$
          \\
\hline
$\Lambda_b^0\to P_p K^-$
          & $1\,(1)$
          & $\Lambda_b^0\to P_{n} K^0$
          & $0.992\,(0.985)$
          \\
$\Lambda_b^0\to P_{\Lambda}\eta'$
          & $0.027\,(4\times10^{-4})$
          & $\Lambda_b^0\to P_{\Lambda}\eta$
          & $0.145\,(0.084)$
          \\
$\Xi^0_b\to P_{\Sigma^+}K^-$
          & $0.819\,(0.692)$
          & $\Xi^0_b\to P_{\Lambda}\bar K^0$
          & $0.166\,(0.165)$
          \\

$\Xi^0_b\to P_{\Lambda}\eta$
          & $0.200\,(0.108)$
          & $\Xi^0_b\to P_{\Lambda}\eta'$
          & $0.027\,(2\times10^{-5})$
          \\
$\Xi^-_b\to P_{\Xi^-}\eta'$
          & $0.025\,(3\times10^{-6})$
          & $\Xi^-_b\to P_{\Xi^-}\eta$
          & $0.196\,(0.104)$
          \\

$\Xi^-_b\to P_{\Sigma^-} \bar K^0$
          & $0.800\,(0.662)$
          & $\Xi^-_b\to P_{\Sigma^0} K^-$
          & $0.168\,(0.168)$
          \\
$\Xi^-_b\to P_{\Lambda} K^-$
          & $0.408\,(0.343)$
          & $\Omega_b^-\to P_{\Xi^-} \bar K^0$
          & $1.15\left|\frac{t_1}{T_1}\right|^2\,\left(1.28\left|\frac{t_1}{T_1}\right|^2\right)$
          \\
$\Omega_b^-\to P_{\Xi^0} K^-$
          & $1.17\left|\frac{t_1}{T_1}\right|^2\,\left(1.33\left|\frac{t_1}{T_1}\right|^2\right)$
          & $\Omega_b^-\to P_{\Xi^{*-}} \bar K^0$
          & $0.209\left|\frac{\tilde t_1}{T_1}\right|^2\,\left(0.139\left|\frac{\tilde t_1}{T_1}\right|^2\right)$
          \\
 $\Omega_b^-\to P_{\Xi^{*0}} K^-$
          & $0.212\left|\frac{\tilde t_1}{T_1}\right|^2\,\left(0.144\left|\frac{\tilde t_1}{T_1}\right|^2\right)$
          & $\Omega_b^-\to P_{\Omega^{-}} \eta$
          & $0.132\left|\frac{\tilde t_1}{T_1}\right|^2\,\left(0.048\left|\frac{\tilde t_1}{T_1}\right|^2\right)$
           \\
%$\Omega_b^-\to P_{\Omega^{-}} \eta'$
%          & $0\,\left(0\right)$
%          &
%          &
%          \\
\hline
$\Lambda_b^0\to P_{\Lambda} K^0$
          & $0.021\,(0.013)$
          & $\Lambda_b^0\to P_n \pi^0$
          & $0.034\,(0.042)$
          \\
$\Lambda_b^0\to P_n\eta$
          & $0.015\,(0.014)$
          & $\Lambda_b^0\to P_n\eta'$
          & $0.004\,(0.001)$
          \\
$\Lambda_b^0\to P_p\pi^-$
          & $0.068\,(0.084)$
          & $\Xi^0_b\to P_{\Xi^0}K^0$
          & $0.029\,(0.017)$
          \\
$\Xi^0_b\to P_{\Lambda}\pi^0$
          & $0.006\,(0.007)$
          & $\Xi^0_b\to P_{\Lambda}\eta$
          & $0.003\,(0.002)$
          \\
$\Xi^0_b\to P_{\Lambda}\eta'$
          & $6\times10^{-4}\,(2\times10^{-4})$
          & $\Xi^0_b\to P_{\Sigma^+}\pi^-$
          & $0.058\,(0.063)$
          \\
$\Xi^0_b\to P_{\Sigma^0}\pi^0$
          & $0.014\,(0.016)$
          &$\Xi^-_b\to P_{\Sigma^0} \pi^-$
          & $0.029\,(0.031)$
           \\
$\Xi^-_b\to P_{\Xi^-} K^0$
          & $0.029\,(0.017)$
          & $\Xi^-_b\to P_{\Sigma^-} \pi^0$
          & $0.028\,(0.031)$
          \\
$\Xi^-_b\to P_{\Sigma^-} \eta$
          & $0.012\,(0.009)$
          & $\Xi^-_b\to P_{\Sigma^-} \eta'$
          & $0.002\,(3\times10^{-4})$
          \\
$\Xi^-_b\to P_{\Lambda} \pi^-$
          & $0.011\,(0.014)$
          & $\Omega_b^-\to P_{\Xi^0} \pi^-$
          &  $0.078\left|\frac{t_1}{T_1}\right|^2\,\left(0.107\left|\frac{t_1}{T_1}\right|^2\right)$
          \\
$\Omega_b^-\to P_{\Xi^-} \eta$
          & $0.017\left|\frac{t_1}{T_1}\right|^2\,\left(0.018\left|\frac{t_1}{T_1}\right|^2\right)$
          & $\Omega_b^-\to P_{\Xi^-} \eta'$
          & $0.005\left|\frac{t_1}{T_1}\right|^2\,\left(0.002\left|\frac{t_1}{T_1}\right|^2\right)$
          \\
$\Omega_b^-\to P_{\Xi^-} \pi^0$
          & $0.038\left|\frac{t_1}{T_1}\right|^2\,\left(0.052\left|\frac{t'_1}{T_1}\right|^2\right)$
          & $\Omega_b^-\to P_{\Omega^-} K^0$
          & $0.020\left|\frac{\tilde t_1}{T_1}\right|^2\,\left(0.008\left|\frac{\tilde t_1}{T_1}\right|^2\right)$
          \\
$\Omega_b^-\to P_{\Xi^{*-}} \pi^0$
          &  $0.008\left|\frac{\tilde t_1}{T_1}\right|^2\,\left(0.007\left|\frac{\tilde t_1}{T_1}\right|^2\right)$
          & $\Omega_b^-\to P_{\Xi^{*0}} \pi^-$
          & $0.016\left|\frac{\tilde t_1}{T_1}\right|^2\,\left(0.015\left|\frac{\tilde t_1}{T_1}\right|^2\right)$
          \\
$\Omega_b^-\to P_{\Xi^{*-}} \eta$
          & $0.003\left|\frac{\tilde t_1}{T_1}\right|^2\,\left(0.002\left|\frac{\tilde t_1}{T_1}\right|^2\right)$
           &$\Omega_b^-\to P_{\Xi^{*-}} \eta'$
          & $3\times10^{-4}\left|\frac{\tilde t_1}{T_1}\right|^2\,\left(7\times10^{-6}\left|\frac{\tilde t_1}{T_1}\right|^2\right)$
\end{tabular}
\end{ruledtabular}
\end{table}
}
%%%%%%%%%%%%%%%%%%%%%%%%%%%%%%%%%%%%%%%%%%%%%%%%%%%%%%%%%%%%%%

In order to extract the branching fraction of $\Lambda_b^0\to P_p^+K^-$ from
the measured branching fraction product Eq. (\ref{eq:BR}), we need to know the decay rate of $P_p^+\to J/\psi p$. The study of the strong decays of pentaquarks is a difficult and yet important task.
For $P_p(4500)^+$, it can decay into $\chi_{c1}p$, $\Sigma_c^{(*)+} \bar D^0$, $\Lambda_c^+ \bar D^{(*)0}$, $J/\psi p$, $\eta_c p$, $\cdots$, etc. A recent work in \cite{Meissner:2015mza} suggests that $\Gamma(P_p(4500)^+)$ is dominated by the $\chi_{c1}p$ channel in spite of its sever phase-space suppression and that $\Br(P_p(4500)^+\to J\psi p)$ is of order 14\% or 24\% depending on the solution for the coupling to $\chi_{c1}p$ and $J/\psi p$.
In \cite{Cheng:2004ew} we have studied the strong decays of light and heavy pentaquarks using the light-front quark model. Along the same line, we plan to investigate the hidden-charm pentaquark strong decays in the forthcoming work. 

\section{Conclusions}

Assuming SU(3) flavor symmetry, we have decomposed the decay amplitudes of bottom baryon decays to a pseudoscalar meson and an octet (a decuplet) pentaquark in terms of three (two) invariant amplitudes $T_1$ and $T_{2,3}$ ($\tilde T_1$ and $\tilde T_2$) corresponding to external $W$-emission and internal $W$-emission diagrams, respectively. For antitriplet bottom baryons $\Lambda_b^0,\Xi_b^0$ and $\Xi_b^-$, their decays to a decuplet pentaquark proceed only through the internal $W$-emission diagram (i.e. $\tilde T_1$ vanishes). On the contrary, the $\Omega_b^-$ decays to the pentaquark (octet or decuplet) can proceed through the external $W$-emission process.
Assuming the dominance from the external $W$-emission amplitudes, we present an estimate of the decay rates relative to $\Lambda_b^0\to P_p^+K^-$. Hence our numerical results will provide a very useful guideline to the experimental search for pentaquarks in bottom baryon decays. For example, $\Xi_b^0\to P_{\Sigma^+}K^-$, $\Xi_b^-\to P_{\Sigma^-}\bar K^0$, $\Omega_b^-\to P_{\Xi^-}\bar K^0$ and $\Omega_b^-\to P_{\Xi^0}K^-$ may have rates comparable to that of $\Lambda_b^0\to P_p^+K^-$ and these modes should be given the higher priority in the experimental searches for pentaquarks.

\section{Acknowledgments}

This research was supported in part by the Ministry of Science and Technology of R.O.C. under Grant
Nos. 103-2112-M-001-005 and  103-2112-M-033-002-MY3.
%, and the National Science Council of R.O.C. under Grant No. NSC100-2112-M-033-001-MY3.

%%%%%%%%%%%%%%%%%%%%%%%%%%%


\begin{thebibliography}{99}

\bibitem{LHCb:penta}
  R.~Aaij {\it et al.} [LHCb Collaboration],
  %``Observation of J/£rp Resonances Consistent with Pentaquark States in £N$_b^0$ ¡÷ J/£rK$^-$p Decays,''
  Phys.\ Rev.\ Lett.\  {\bf 115}, 072001 (2015)
  [arXiv:1507.03414 [hep-ex]].

\bibitem{LHCb:Br}
  R.~Aaij {\it et al.} [LHCb Collaboration],
  %``Study of the production of $\Lambda_b^0$ and $\overline{B}^0$ hadrons in $pp$ collisions and first measurement of the $\Lambda_b^0\rightarrow J/\psi pK^-$ branching fraction,''
  arXiv:1509.00292 [hep-ex].

\bibitem{Maiani:2015vwa}
  L.~Maiani, A.~D.~Polosa and V.~Riquer,
  %``The New Pentaquarks in the Diquark Model,''
  Phys.\ Lett.\ B {\bf 749}, 289 (2015)
  [arXiv:1507.04980 [hep-ph]].

\bibitem{Anisovich:2015cia}
  V.~V.~Anisovich, M.~A.~Matveev, J.~Nyiri, A.~V.~Sarantsev and A.~N.~Semenova,
  %``Pentaquarks and resonances in the $pJ/\psi$ spectrum,''
  arXiv:1507.07652 [hep-ph].

\bibitem{Wang:2015epa}
  Z.~G.~Wang,
  %``Analysis of the $P_c(4380)$ and $P_c(4450)$ as pentaquark states in the diquark model with QCD sum rules,''
  arXiv:1508.01468 [hep-ph];
  Z.~G.~Wang and T.~Huang,
  %``Analysis of the ${\frac{1}{2}}^{\pm}$ pentaquark states in the diquark model with QCD sum rules,''
  arXiv:1508.04189 [hep-ph].

\bibitem{Jaffe:2003sg}
  R.~L.~Jaffe and F.~Wilczek,
  %``Diquarks and exotic spectroscopy,''
  Phys.\ Rev.\ Lett.\  {\bf 91}, 232003 (2003)
  [hep-ph/0307341].

\bibitem{Lebed:2015tna}
  R.~F.~Lebed,
  %``The Pentaquark Candidates in the Dynamical Diquark Picture,''
  Phys.\ Lett.\ B {\bf 749}, 454 (2015)
  [arXiv:1507.05867 [hep-ph]].

\bibitem{Karliner:2004qw}
  M.~Karliner and H.~J.~Lipkin,
  %``The Narrow width of the Theta+: A Possible explanation,''
  Phys.\ Lett.\ B {\bf 586}, 303 (2004)
  [hep-ph/0401072].

\bibitem{Chen:2015loa}
  R.~Chen, X.~Liu, X.~Q.~Li and S.~L.~Zhu,
  %``Identifying exotic hidden-charm pentaquarks,''
  arXiv:1507.03704 [hep-ph].

\bibitem{Chen:2015moa}
  H.~X.~Chen, W.~Chen, X.~Liu, T.~G.~Steele and S.~L.~Zhu,
  %``Towards exotic hidden-charm pentaquarks in QCD,''
  arXiv:1507.03717 [hep-ph].

\bibitem{Roca:2015dva}
  L.~Roca, J.~Nieves and E.~Oset,
  %``The LHCb pentaquark as a $\bar{D}^*\Sigma_c-\bar{D}^*\Sigma_c^*$ molecular state,''
  arXiv:1507.04249 [hep-ph].

\bibitem{He:2015cea}
  J.~He,
  %``The $\bar{D}\Sigma^*_c$ and $\bar{D}^*\Sigma_c$ interactions and the LHCb hidden-charmed pentaquarks,''
  arXiv:1507.05200 [hep-ph].

\bibitem{Burns:2015dwa}
  T.~J.~Burns,
  %``Phenomenology of $P_c(4380)^+$, $P_c(4450)^+$ and related states,''
  arXiv:1509.02460 [hep-ph].

\bibitem{Meissner:2015mza}
  U.~G.~Meisner and J.~A.~Oller,
  %``Testing the $\chi_{c1}\, p$ composite nature of the $P_c(4450)$,''
  arXiv:1507.07478 [hep-ph].

\bibitem{Kubarovsky:2015aaa}
  V.~Kubarovsky and M.~B.~Voloshin,
  %``Formation of hidden-charm pentaquarks in photon-nucleon collisions,''
  Phys.\ Rev.\ D {\bf 92}, 031502 (2015)
  [arXiv:1508.00888 [hep-ph]].

\bibitem{Scoccola:2015nia}
  N.~N.~Scoccola, D.~O.~Riska and M.~Rho,
  %``On the pentaquark candidates $P_c^+(4380)$ and $P_c^+(4450)$ within the soliton picture of baryons,''
  arXiv:1508.01172 [hep-ph].

\bibitem{Guo:2015umn}
  F.~K.~Guo, U.~G.~Meisner, W.~Wang and Z.~Yang,
  %``How to reveal the exotic nature of the P_c(4450),''
  arXiv:1507.04950 [hep-ph].

\bibitem{Liu:2015fea}
  X.~H.~Liu, Q.~Wang and Q.~Zhao,
  %``Understanding the newly observed heavy pentaquark candidates,''
  arXiv:1507.05359 [hep-ph].

\bibitem{Mikhasenko:2015vca}
  M.~Mikhasenko,
  %``A triangle singularity and the LHCb pentaquarks,''
  arXiv:1507.06552 [hep-ph].

\bibitem{Anisovich}
  V.~V.~Anisovich, M.~A.~Matveev, J.~Nyiri, A.~V.~Sarantsev and A.~N.~Semenova,
  arXiv:1509.03028 [hep-ph].

\bibitem{Wang:2015jsa}
  Q.~Wang, X.~H.~Liu and Q.~Zhao,
  %``Photoproduction of hidden charm pentaquark states $P_c^+(4380)$ and $P_c^+(4450)$,''
  Phys.\ Rev.\ D {\bf 92}, 034022 (2015)
  [arXiv:1508.00339 [hep-ph]].

\bibitem{Karliner:2015voa}
  M.~Karliner and J.~L.~Rosner,
  %``Photoproduction of Exotic Baryon Resonances,''
  arXiv:1508.01496 [hep-ph].

\bibitem{Chua:2003it}
  C.~K.~Chua,
  %``Charmless two body baryonic B decays,''
  Phys.\ Rev.\ D {\bf 68}, 074001 (2003)
  [hep-ph/0306092].
  %%CITATION = HEP-PH/0306092;%%
  %11 citations counted in INSPIRE as of 08 Dec 2014

\bibitem{Chua:2013zga}
  C.~K.~Chua,
  %``Charmless Two-body Baryonic $B_{u,d,s}$ Decays Revisited,''
  Phys.\ Rev.\ D {\bf 89}, 056003 (2014)
  [arXiv:1312.2335 [hep-ph]].
  %%CITATION = ARXIV:1312.2335;%%
  %2 citations counted in INSPIRE as of 08 Dec 2014

\bibitem{KLOE}
  F.~Ambrosino {\it et al.},
  %``A global fit to determine the pseudoscalar mixing angle and the gluonium
  %content of the eta' meson,''
  JHEP {\bf 0907}, 105 (2009)
  [arXiv:0906.3819 [hep-ph]].

\bibitem{FKS} T. Feldmann, P. Kroll, and B. Stech, Phys. Rev. D {\bf 58}, 114006 (1998); \pl {\bf B449}, 339 (1999).

\bibitem{Li:2015gta}
  G.~N.~Li, M.~He and X.~G.~He,
  %``Some Predictions of Diquark Model for Hidden Charm Pentaquark Discovered at the LHCb,''
  arXiv:1507.08252 [hep-ph].

%\bibitem{PDG}
 % K.A. Olive {\it et al.} (Particle Data Group), Chin. Phys. C {\bf 38}, 090001 (2014).

\bibitem{Hsiao:2015nna}
  Y.~K.~Hsiao and C.~Q.~Geng,
  %``Pentaquarks from intrinsic charms in $\Lambda_b$ decays,''
  arXiv:1508.03910 [hep-ph].

\bibitem{Cheng:2004ew}
  H.~Y.~Cheng and C.~K.~Chua,
  %``Light-front approach for Pentaquark strong decays,''
  JHEP {\bf 0411}, 072 (2004)
  [hep-ph/0406036].
  
\end{thebibliography}
\end{document}